\documentclass[twocolumn,amsmath,amssymb,superscriptaddress]{revtex4-2}

\usepackage{graphicx}
\usepackage{color}
\usepackage{multirow}
\usepackage[normalem]{ulem}

\begin{document}

\title{The de Sitter swampland conjectures in the context of Chaplygin-inspired inflation}

\author{Orfeu Bertolami}
\email{orfeu.bertolami@fc.up.pt}
\affiliation{Departamento de F\'{\i}sica e Astronomia, Faculdade de Ci\^encias, Universidade do Porto, Rua do Campo Alegre s/n, 4169-007 Porto, Portugal\\}
\affiliation{Centro de F\'{\i}sica das Universidades do  Minho e do Porto, Rua do Campo Alegre s/n, 4169-007 Porto, Portugal}

\author{Robertus Potting}
\email{rpotting@ualg.pt}
\affiliation{Departamento de F\'{\i}sica, Faculdade de Ci\^encias e Tecnologia, Universidade do Algarve, Campus de Gambelas, 8005-139 Faro, Portugal\\}
\affiliation{CENTRA, Instituto Superior T\'ecnico, Universidade de Lisboa, Avenida Rovisco Pais, 1049-001 Lisboa, Portugal}

\author{Paulo M. S\'a}
\email{pmsa@ualg.pt}
\affiliation{Departamento de F\'{\i}sica, Faculdade de Ci\^encias e Tecnologia, Universidade do Algarve, Campus de Gambelas, 8005-139 Faro, Portugal\\}
\affiliation{Instituto de Astrof\'{\i}sica e Ci\^encias do Espa\c co, Faculdade de Ci\^encias, Universidade de Lisboa, Campo Grande, 1749-016 Lisboa, Portugal}

\begin{abstract}
In this work, we discuss the de Sitter swampland conjectures in the context of the generalized Chaplygin-inspired inflationary model. We demonstrate that these conjectures can be satisfied, but only in the region of the parameter space far away from the General Relativity limit. The cosmic microwave background data had already been found to restrict the allowed inflationary potentials of this model. Our results impose a further limitation on the possible potentials.
\end{abstract}

\maketitle

Swampland conjectures have been put forward in order to identify de Sitter solutions that do not lie in the string theory landscape.
Such apparently consistent solutions do not admit a suitable ultraviolet completion (see Ref.~\cite{Vafa} for the original discussion and Ref.~\cite{Palti} for a review).
These conjectures are particularly relevant as they relate the intrinsic consistency of string theory to conditions for obtaining our four-dimensional world, and also address the notoriously difficult problem of obtaining inflation from the fields that naturally arise in string theory. 

In fact, quite involved scenarios are required in string theory in order to get inflation (see, for instance, Ref.~\cite{KKLT}), which is somewhat surprising as in $N=1$ supergravity, presumably a low-energy limit of string theory, inflationary solutions can be naturally implemented (see eg.~Ref.~\cite{Adams}).
From a phenomenological point of view, viable string theory models, namely those that have an intermediate-scale Grand Unified Theory, have been shown to require a period of inflation for their implementation \cite{OBRoss87}.

In broad terms, the swampland conjectures amount to a set of necessary conditions that ensure general low-energy features, such as the presence of local gauge symmetries, as well as the presence of at least one Planck-mass particle in order to account for the weakness of gravity.
They are also required in order to assure that higher-derivative terms in the effective action do not lead to superluminal propagation \cite{Vafa3}.
Not included among these general requirements is the Strong Equivalence Principle, from which it follows that gravity has to be described by General Relativity.
Nevertheless, in most of the applications of the swampland conjectures, this latter assumption is tacitly made.
Technically, this implies that when considering the more general setting of alternative theories of gravity, they are confronted with the swampland conjectures in the so-called Einstein frame.  

Indeed, given that General Relativity has some limitations, it is natural to consider alternative theories of gravity \cite{Clifton} and ask if the swampland conjectures hold for inflationary models (either single-field or more
involved) arising from these theories.
This issue has been recently analyzed for theories of gravity with non-minimal coupling between curvature and matter \cite{BBHL}, and their inflationary solutions \cite{GRB}.
It was found that inflation in these theories cannot be reconciled with the de Sitter swampland conjectures \cite{BGS2023}.

In the current work, we examine another model in the context of the swampland conjectures, namely the generalized Chaplygin-inspired single field inflationary model (see Ref.~\cite{OBD} for details).
This confrontation is particularly interesting, as it allows, through inflation, to examine models that go beyond the Standard Model of particle physics.
An additional motivation to consider the generalized Chaplygin-inspired inflationary model concerns the recent claim that a Chaplygin-like equation of state can endow the vacuum with interesting features related to entanglement entropy generation \cite{OB23}.

The swampland conjectures can be expressed through constraints on scalar fields in low-energy effective field theories, generically denoted by $\phi$ \cite{OV,OOSV}, namely (in the Einstein frame),
\begin{gather}
\label{eq:SC1}
\frac{\Delta \phi}{M_{\rm P}} < c_1,
\\
M_{\rm P}\frac{|V'|}{V} > c_2,
\label{eq:SC2}
\end{gather}
where $\Delta\phi$ denotes a scalar-field variation, $M_{\rm P}\equiv M_{\rm Pl}/\sqrt{8\pi}$ is the reduced Planck's mass, $V(\phi)$ is the scalar field potential, $V' \equiv \partial V/ \partial\phi$, and $c_1$ and $c_2$ are $\mathcal{O}(1)$ constants.
It has been argued \cite{OPSV,garg-krishnan} that a more refined condition
should also be considered, namely the potential should satisfy condition (\ref{eq:SC2}), or, alternatively,
\begin{equation}
M_{\rm P}^2\frac{V''}{V} < -c_3,
\label{eq:SC3}
\end{equation}
where $V'' \equiv \partial^2 V/\partial\phi^2$ and the constant $c_3$ is of order one.

The requirements expressed by Eqs.~(\ref{eq:SC2}) and (\ref{eq:SC3}) are incompatible with the onset conditions of single-field (cold) inflation, which require that the inflaton satisfies the slow-roll conditions $\epsilon_{\phi} \ll 1$ and $|\eta_{\phi}| \ll 1$ at the onset of inflation \cite{PDG2022}, where
\begin{equation}
\epsilon_{\phi}= \frac{M_{\rm P}^2}{2} \left(\frac{V'}{V} \right)^2
\label{eq:epsilon}
\end{equation}
and
\begin{equation}
\eta_{\phi}=M_{\rm P}^2  \frac{V''}{ V},
\label{eq:eta}
\end{equation}
so that at the end of inflation $\epsilon_{\phi} \sim |\eta_{\phi}| \sim1$.
It is remarkable that the conditions for the onset of inflation match the observational results \cite{PDG2022}
\begin{equation}
\epsilon_{\phi} <0.0044
\label{eq:ePlanck}
\end{equation}
and
\begin{equation}
\eta_{\phi} = - 0.015 \pm 0.006,
\label{eq:etaPlanck}
\end{equation}
which are incompatible with the requirements on $c_2$ and $c_3$. 

However, this incompatibility can be alleviated or resolved if one considers multi-field inflationary backgrounds which follow curved, non-geodesic trajectories in field space \cite{Achucarro} and also if one considers either excited initial states for tensor perturbations \cite{Ashoorioon}, chaotic inflation on the brane \cite{Lin}, or a significant dissipation in the context of warm inflationary models \cite{Berera} for one \cite{mkr,bkr} or any number of scalar fields \cite{BS22}.

The generalized Chaplygin gas model has been originally proposed  to unify dark matter and dark energy by resorting to a fluid with an exotic equation of state
\begin{equation}
p = - \frac{A}{\rho^{\alpha}},
\label{eq:eqstate}
\end{equation}
where $p$ is the isotropic pressure, $\rho$ is the energy density, $A$ a positive constant, and $0< \alpha \leq 1$.
The original Chaplygin gas model corresponds to $\alpha=1$.
It is well known that this equation of state has many interesting features \cite{Kamen,Bilic,BBS}.
The support of the Chaplygin equation of state $(\alpha=1)$ is a Born-Infeld action that describes a brane \cite{Bilic} which can be parametrised through the action of a complex \cite{Bilic,BBS} or a real scalar field \cite{Kamen,BSST}.

The evolution equation for the energy density,
\begin{equation}
\dot{\rho} + 3 H (\rho + p)  = 0,
\label{eq:conservation}
\end{equation}
where the dot denotes the derivative with respect to the cosmic time, $H = \dot{a}/a$ is the expansion rate and $a(t)$ is the scale factor of the Friedmann-Lema\^itre-Robertson-Walker metric, with $p$ given by Eq.~(\ref{eq:eqstate}), can be easily integrated, yielding \cite{Kamen,Bilic,BBS}
\begin{equation}
\rho = \left[A + \frac{B}{a^{3(1 + \alpha)}}
\right]^{\frac{1}{1+ \alpha}},
\label{eq:rho}
\end{equation}
where $B$ is an integration constant. 

This type of behavior of the energy density $\rho$ can also arise from a modification of gravity, particularly from a generalized Born-Infeld action for a scalar field $\phi$ with energy density $\rho_{\phi}$, giving rise to a modified Friedmann equation of the form \cite{OBD}
\begin{equation}
	H^2 = \frac{1}{3  M_{\rm P}^2}\left[A + \rho_{\phi}^{1+\alpha}
	\right]^{\frac{1}{1+ \alpha}},
\label{eq:H}
\end{equation}
where the scalar field satisfies the usual equation of motion
\begin{equation}
\ddot{\phi} + 3 H \dot{\phi} + V' = 0,
\label{eq:phi}
\end{equation}
with $V$ a suitable inflationary potential. Note that the setup proposed by Eq.\ (\ref{eq:H}) differs from the one where the Chaplygin gas energy density expression, Eq.\ (\ref{eq:rho}), is considered in a braneworld scenario. Its adequacy with respect to the de Sitter swampland conditions was discussed in Ref.\ \cite{MJ} for warm inflation.
The Chaplygin-inspired model suggested in Ref.\ \cite{OBD} and discussed here assumes a change of gravity itself that amounts to a modification of the standard Friedmann equation as in Eq.\ (\ref{eq:H}), in which the contribution of the energy density of matter in Eq. (\ref{eq:rho}) is replaced by the energy density of the inflaton.

Compatibility with the cosmic microwave background (CMB) data for monomial and hilltop potentials has been examined for the generalized Chaplygin-inspired inflationary model described by Eqs.~(\ref{eq:H}) and (\ref{eq:phi}) with respect to Planck data in the $r$-$n_s$ plane \cite{GBR}.
Interestingly, compatibility with CMB Planck data is ensured in the limit $A \gg V^{1+\alpha}$ for hilltop potentials, a situation where the inflationary features of the generalized Chaplygin-inspired model differ from the ones arising from the General Relativity limit, $A \ll V^{1+\alpha}$, where the expansion rate is given by the usual Friedmann equation.
On the other hand, monomial models are not compatible with CMB data for $A \gg V^{1+\alpha}$.
We shall see below that these results can be further restricted by the swampland conjectures (\ref{eq:SC2}) and (\ref{eq:SC3}). 

Now let us investigate if the generalized Chaplygin-inspired inflationary model described by Eqs.~(\ref{eq:H}) and (\ref{eq:phi}) satisfies the de Sitter swampland conjectures (\ref{eq:SC2}) and (\ref{eq:SC3}) for a generic form of the potential $V(\phi)$ and an arbitrary nonvanishing value of the constant $A$.

From Eqs.~(\ref{eq:H}) and (\ref{eq:phi}) it is straightforward to obtain an expression for the time derivative of the Hubble parameter, namely,
\begin{equation}
	\dot{H}=-\frac{1}{2M_{\rm P}^2} 
	\rho_\phi^\alpha (\rho_\phi+p_\phi)
	\left( A+\rho_\phi^{1+\alpha}\right)^{-\frac{\alpha}{1+\alpha}},
	\label{eq: dotH}
\end{equation}
where $\rho_\phi=\dot{\phi}^2/2+V$ and $p_\phi=\dot{\phi}^2/2-V$ are the energy density and pressure of the scalar field $\phi$, respectively.

In the slow-roll approximation, for which $\dot{\phi}^2/2\ll V$ and $|\ddot{\phi}|\ll|3H\dot{\phi}|$, Eqs.~(\ref{eq:H}), (\ref{eq:phi}), and (\ref{eq: dotH}) can be written as
\begin{gather}
 	H^2 \simeq \frac{1}{3M_{\rm P}^2} 
 	\left( A+V^{1+\alpha} \right)^\frac{1}{1+\alpha}, \label{H}
 	\\
 	\dot{\phi}  \simeq - \frac{M_{\rm P} V'}{\sqrt3}
 	\left( A+V^{1+\alpha} \right)^{-\frac{1}{2(1+\alpha)}}, \label{phidot}
 	\\
 	\dot{H} \simeq -\frac{1}{6} V^\alpha (V')^2
	\left( A+V^{1+\alpha}\right)^{-1}, \label{Hdot}
\end{gather}
where the symbol $\simeq$ means ``equal within the slow-roll approximation".

Using the above equations, the quantities $-\dot{H}/H^2$ and $\ddot{\phi}/(H\dot{\phi})$ can be related to the slow-roll parameters $\epsilon_\phi$ and $\eta_\phi$ defined in  Eqs.~(\ref{eq:epsilon}) and (\ref{eq:eta}), respectively.
We obtain
\begin{gather}
-\frac{\dot{H}}{H^2} \simeq \epsilon_\phi \left( 1+\frac{A}{V^{1+\alpha}} 
\right)^{-\frac{2+\alpha}{1+\alpha}},
\\
\frac{\ddot{\phi}}{H\dot{\phi}} \simeq \epsilon_\phi 
\left( 1+\frac{A}{V^{1+\alpha}} \right)^{-\frac{2+\alpha}{1+\alpha}} 
- \eta_\phi \left( 1+\frac{A}{V^{1+\alpha}} 
\right)^{-\frac{1}{1+\alpha}}.
\end{gather}

Taking into account that in the slow-roll regime $|\dot{H}|/H^2 \allowbreak \ll1$ and $|\ddot{\phi}/(H\dot{\phi})|\ll 1$, we conclude that
\begin{gather}
	\epsilon_\phi \ll \left( 1+\frac{A}{V^{1+\alpha}} 
	\right)^{\frac{2+\alpha}{1+\alpha}},
	\label{epsilon}
\\
	|\eta_\phi| \ll \left( 1+\frac{A}{V^{1+\alpha}} 
\right)^{\frac{1}{1+\alpha}}.
	\label{eta}
\end{gather}
Note that the parameters $\epsilon_\phi$ and $\eta_\phi$ are related to the constants $c_2$ and $c_3$ of the de Sitter swampland conjectures [see Eqs.~(\ref{eq:SC2}) and (\ref{eq:SC3})] through the relations
\begin{equation}
	c_2^2 < 2\epsilon_\phi \quad \mbox{and} \quad c_3 < |\eta_\phi|.
	\label{c2c3}
\end{equation}

In the regime $A\ll V^{1+\alpha}$, Eqs.~(\ref{epsilon}) and (\ref{eta}) reduce to $\epsilon_\phi\ll1$ and $|\eta_\phi| \ll1$, implying, in turn, that $c_2\ll1$ and $c_3\ll1$, which goes against the de Sitter swampland conjectures.
This result coincides with that of General Relativity for the single-field (cold) inflation case.

Much more interesting is the regime $A\gg V^{1+\alpha}$.
In this case, Eqs.~(\ref{epsilon}) and (\ref{eta}) reduce to
\begin{equation}
	\epsilon_\phi \ll \left(\frac{A}{V^{1+\alpha}} 
	\right)^{\frac{2+\alpha}{1+\alpha}}
	\quad \mbox{and} \quad
	|\eta_\phi| \ll \left( \frac{A}{V^{1+\alpha}} 
\right)^{\frac{1}{1+\alpha}},
\end{equation}
which allow $\epsilon_\phi$ and $|\eta_\phi|$ to take values larger that one.
Relations~(\ref{c2c3}) then imply that both $c_2$ and $c_3$ can be of order one during quasi-exponential inflation, thus satisfying de Sitter swampland conjectures.

Let us now turn to the swampland distance conjecture, given by Eq.~(\ref{eq:SC1}).
The number of e-foldings of inflation is given by
\begin{equation}
 N=\ln \left( \frac{a_f}{a_i} \right)=\int_{a_i}^{a_f} \! \frac{da}{a}
 = \int_{\phi_i}^{\phi_f} \! H\frac{d\phi}{\dot{\phi}},
\end{equation}
where the subscripts $i$ and $f$ denote the inflationary period's beginning and end, respectively, and typically $N\sim50-60$.
In the slow-roll regime, the above expression can be approximated by
\begin{equation}
N \simeq \frac{H}{|\dot{\phi}|} |\Delta\phi|.
\end{equation}
Now, using Eqs.~(\ref{H}) and (\ref{phidot}), the field excursion for the inflaton can be written as
\begin{equation}
 |\Delta\phi| \simeq N M_{\rm P}^2 \left| \frac{V'}{V} \right|
\left( 1+\dfrac{A}{V^{1+\alpha}} \right)^{-\frac{1}{1+\alpha}}
\end{equation}
or, taking into account the definition of $\epsilon_\phi$ given by Eq.~(\ref{eq:epsilon}), as
\begin{equation}
|\Delta\phi| \simeq N M_{\rm P} \sqrt{2\epsilon_\phi}
\left( 1+\dfrac{A}{V^{1+\alpha}} \right)^{-\frac{1}{1+\alpha}}.
\end{equation}
Finally, using Eq.~(\ref{epsilon}), we obtain
\begin{equation}
\frac{|\Delta\phi|}{M_{\rm P}} \ll \sqrt2 N
\left( 1+\frac{A}{V^{1+\alpha}} \right)^{\frac{\alpha}{2(1+\alpha)}},
\end{equation}
from which follows that the swampland distance conjecture, given by Eq.~(1), can be satisfied in both regimes, $A/V^{1+\alpha}\ll1$ or $A/V^{1+\alpha}\gg1$.

Compatibility of the generalized Chaplygin-inspired mo\-del with CMB data has been investigated in Ref.~\cite{GBR}.
It was found that for potentials of the form $V(\phi) = V_0(\phi/M_p)^n$ with $n = 1, 2$, compatibility with CMB data is not possible for $x \equiv \frac{A}{V^{1+\alpha}} > 1$.
Hence, the latter are incompatible with the swampland conjectures (\ref{eq:SC2}) and (\ref{eq:SC3}).

On the other hand, for quadratic and quartic hilltop potential models, $V(\phi) = V_0\bigl(1- \frac{\gamma}{n} (\phi/M_p)^n\bigr)$ with $0 < \gamma < 1$ and $n = 2, 4$, compatibility with CMB data was explicitly shown to be possible for $\alpha = 0.5$ or $1$ for a wide range (including large) values of $x$.
Therefore, these potentials can be made compatible with the swampland conjectures  (\ref{eq:SC1}), (\ref{eq:SC2}), and (\ref{eq:SC3}).

In conclusion, in this work we have considered the generalized Chaplygin-inspired single field inflationary model and shown that it is compatible with the swampland conjectures (\ref{eq:SC2}) and (\ref{eq:SC3}) provided the condition $x \gg 1$ is satisfied.
Interestingly, this limit corresponds to the situation where some of the features of the model differ from the ones arising in the General Relativity limit $x \ll 1$.
As we have seen, compatibility of the generalized Chaplygin-inspired inflationary model with the CMB data restricted the allowed potentials.
The de Sitter swampland conjectures impose a further limitation on the possible potentials.
In fact, these conjectures suggest that one should stretch the model parameter
$x$ to large values, far from the General Relativity limit.

\newpage

\begin{acknowledgments}
We gratefully acknowledge support from the Funda\c{c}\~ao para 
a Ci\^encia e a Tecnologia (Portugal) through the research grants
doi.org/10.54499/CERN/FIS-PAR/0027/ 2021 (OB),
doi.org/10.54499/UIDB/00099/2020 (RP),
doi.org/10.54499/UIDB/04434/2020, and doi.org/ 10.54499/UIDP/04434/2020 (PMS).
\end{acknowledgments}

\end{document}